\documentstyle[epsfig,psfig,12pt]{article} 
\def\ra{\rightarrow}
\def\slash#1{#1\!\!\! /}
\def\beq{\begin{equation}} 
\def\eeq{\end{equation}} 
\def\beqa{\begin{eqnarray}} 
\def\eeqa{\end{eqnarray}} 
\def\ba{\begin{array}} 
\def\ea{\end{array}} 
\def\vb#1{\vbox to #1 pt{}}
\def\ovl{\overline}

\def\ni{\noindent}
\def\bs{\bigskip}
 
\begin{document} 
\begin{titlepage} 
\begin{flushright} 
FISIST/10-98/CFIF\\ 
\end{flushright} 
\vspace*{5mm} 
\begin{center}  
{\Large \bf Vector Boson decays of the Higgs Boson}\\[15mm] 

{\large{Jorge C. Rom\~ao${}^1$ and Sofia Andringa${}^2$ } 
\hspace{3cm}\\
{\small ${}^1$Departamento de F\'\i sica, Instituto Superior T\'ecnico}\\
\vspace{-3.5mm}
{\small A. Rovisco Pais, P-1096 Lisboa Codex, Portugal}\\ 
{\small ${}^2$LIP, Av. Elias Garcia 14, 1e, P-1000 Lisbon, Portugal\\ }}
\end{center}
\vspace{5mm}

\begin{abstract} 
We derive the width of the Higgs boson into vector bosons. General formulas
are derived both for the on--shell decay $H \ra VV$ as well for the 
off--shell decays, $H \ra V^* V$ and $H \ra V^* V^*$, where
$V=\gamma,W^{\pm},Z^0$. For the off-shell decays the width of the decaying
vector boson is properly included. The formulas are valid both for the
Standard Model as well as for arbitrary extensions. As an example we
study in detail the gauge-invariant effective Lagrangian models where
we can have sizable enhancements over the Standard Model that could
be observed at LEP.

\end{abstract} 
 
\end{titlepage} 

\setcounter{page}{1} 

\section{Introduction}

In recent years it has been established~\cite{LEPI} with great precision (in some cases
better than 0.1\%) that the interactions of the gauge bosons with the
fermions are described by the Standard Model (SM)~\cite{SM}. However other
sectors of the SM have been tested to a much lesser degree. In fact only now
we are beginning to probe the self--interactions of the gauge bosons through
their pair production at the Tevatron~\cite{Tevatron} and LEP~\cite{LEPII}
and the Higgs sector, responsible for the symmetry breaking has not yet been
tested. 

\ni
A more complicated symmetry breaking sector can introduce modifications in
the couplings of the Higgs boson with the vector bosons. It is therefore
important to have expressions for the decay widths of the Higgs boson into
vector bosons that are valid for an arbitrary extension of the SM. For the
region of the Higgs boson mass relevant for searches at 
LEP II and LHC it is necessary that the vector bosons in the decays can 
be off--shell.

\ni
In this paper we derive the complete set of formulas for the decay
widths of the Higgs boson in vector bosons. The formulas are valid
both for the Standard Model (SM) and for any arbitrary extension.
For the case of the decay into the $W^\pm$ and $Z^0$ the formulas are
also valid for off--shell decays. This is important for Higgs boson
masses close to the threshold of the production of one or two
real vector bosons. Many of these results have appeared before in the
literature~\cite{marciano,HHG,hagiwara2,hagiwara,ee500,djouadi}, sometimes for
particular cases, but we think
that it will be very useful for the Higgs boson search at
LEP and at LHC to have the general results in a consistent notation.

\ni
The paper is organized as follows. In Section 2 the decays $H\ra VV$
where $V=W^\pm,Z^0$ are calculated. The decays $H\ra \gamma \gamma$
and $H\ra \gamma Z^0$, that in the SM proceed at one--loop level, are
reviewed in Sections 3 and 4, respectively. In Section 5 the off-shell
3--point functions $Z^*\ra H \gamma$ and $\gamma^*\ra H \gamma$ are
given in a consistent notation both for the SM as well as for any of
its extensions. In Section 6 we give an example of physics Beyond de
Standard Model (BSM) and in Section 7 a brief discussion of our
results and a comparison with previous ones is presented.

\section{The Decays $H\ra V V$}
 
\subsection{The $HVV$ Couplings}

We consider the most general couplings of the Higgs $H$ with the
$W^\pm$ and $Z^0$. These are

\vskip 1.5cm
\beq
\hskip 5cm i\, g M_V \left( g_{\mu \nu} +T^V_{\mu \nu} \right)
\eeq

\begin{picture}(0,0.5)
\put(2,-1){\psfig{figure=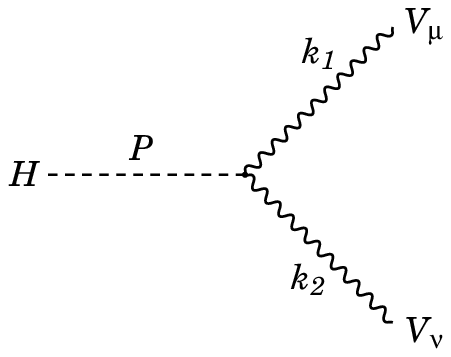}}
\end{picture}

\vskip 0.5cm

\ni
where $V=W,Z$ and $T^W_{\mu \nu}$ and $T^Z_{\mu \nu}$ are the extra 
contributions
from new physics Beyond the Standard Model (BSM). In general they will
depend on the momenta $P$, $k_1$ and $k_2$, but as we will see, we will not
need their exact expressions to get the final formulas. 

\subsection{The On--Shell Decay $H\ra VV$}

We now consider the on--shell decay $H\ra VV$. To be precise we derive
the expression for $H\ra W^+ W^-$ and then present a final result
valid also for $H \ra Z^0 Z^0$. We consider the kinematics given in
Fig. 1.

\vbox{
\begin{picture}(0,4.5)
\put(4.5,0){\psfig{figure=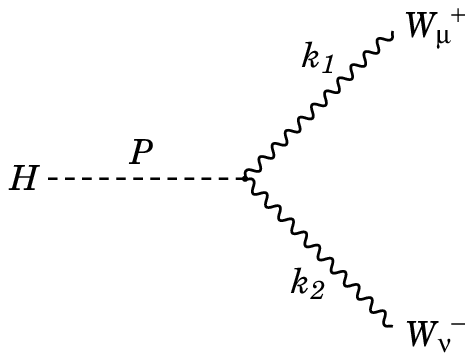}}
\end{picture}

\centerline{Fig. 1}
}

\noindent
The differential width is given by~\cite{PDG}

\beq
d \Gamma= \frac{1}{32 \pi^2}\, \sum_{pol} | {\cal M}|^2\, \frac{|\vec k_1
|}{M_H^2}\, d \Omega_{\vec k_1}
\label{dghvv}
\eeq
where
\beq
{\cal M}= i\, gM_W \epsilon^{\mu}(k_1) \epsilon^{\nu}(k_2)\,
\left( g_{\mu \nu} +T^W_{\mu \nu} \right)
\eeq
We get therefore
\beqa
\sum_{pol} | {\cal M} |^2 &=& (gM_W)^2\, 
\left( - g^{\mu \alpha} +\frac{k_1^{\mu } k_1^{\alpha}}{M_W^2} \right)  
\left( - g^{\nu \beta} +\frac{k_2^{\nu } k_2^{\beta}}{M_W^2} \right)
\left( g_{\mu \nu} +T^W_{\mu \nu} \right)
\left( g_{\alpha \beta} +T^W_{\alpha \beta} \right) \cr
\vb{24}
&=&
\left[
2 + \frac{\left(k_1 \cdot k_2\right)^2}{M_W^4} + 2 T_{\alpha}^{W \alpha} 
- 2 \frac{k_1^{\alpha} k_1^{\beta}}{M_W^2} T^W_{\alpha \beta}
- 2 \frac{k_2^{\alpha} k_2^{\beta}}{M_W^2} T^W_{\alpha \beta}
+ 2 \frac{k_1 \cdot k_2 k_1^{\alpha} k_2^{\beta}}{M_W^4} T^W_{\alpha
\beta} \right.\cr
\vb{24}
&&\hskip 0.5cm \left.
+T^W_{\mu \nu} T^{W \mu \nu}
-\frac{k_{1 \mu}k_1^{\alpha}}{M_W^2} T^{\mu \nu} T^W_{\alpha \nu}
-\frac{k_{2 \nu}k_2^{\beta}}{M_W^2} T^{\alpha \nu} T^W_{\alpha \beta}
+  \frac{k_1^{\mu} k_2^{\nu}}{M_W^2} T^W_{\mu\nu}\,
 \frac{k_1^{\alpha} k_2^{\beta}}{M_W^2} T^W_{\alpha\beta}
\right]
\eeqa

\noindent
Now, using 
\beqa
k_1 \cdot k_2 &=& \frac{1}{2} (M_H^2 -2 M_W^2) \cr
\vb{20}
&=&
\frac{1}{2} \sqrt{M_H^4 \lambda(M_W^2,M_W^2;M_H^2) +4 M_W^4}
\eeqa
where
\beq
\lambda(x,y;z)=\left(1-\frac{x}{z}-\frac{y}{z}\right)^2 
-4\frac{xy}{z^2}
\label{lambdadef}
\eeq
and defining
\beqa
X(p_1,p_2,M_H,T^V)&\equiv&
4 \left[
 2 \frac{p_1^2 p_2^2}{M_H^4}\, T_{\alpha}^{V \alpha} 
- 2 \frac{p_2^2}{M_H^2}\, 
\frac{p_1^{\alpha} p_1^{\beta}}{M_H^2}\, T^V_{\alpha \beta}
- 2 \frac{p_1^2}{M_H^2}\, 
\frac{p_2^{\alpha} p_2^{\beta}}{M_H^2}\, T^V_{\alpha \beta}
\right.\cr
\vb{24}
&&\hskip 0.5cm 
+ 2 \frac{p_1 \cdot p_2 p_1^{\alpha} p_2^{\beta}}{M_H^4}\, T^V_{\alpha\beta} 
+ \frac{p_1^2 p_2^2}{M_H^4} T^V_{\mu \nu}\, T^{V \mu \nu}
- \frac{p_2^2}{M_H^2}\, 
\frac{p_{1 \mu}p_1^{\alpha}}{M_H^2}\, T^{V \mu \nu} T^V_{\alpha \nu}
\cr
\vb{24}
&&\hskip 0.5cm \left.
- \frac{p_1^2}{M_H^2}\, 
\frac{p_{2 \nu}p_2^{\beta}}{M_H^2}\, T^{V \alpha \nu} T^V_{\alpha
\beta}
+  \frac{p_1^{\mu} p_2^{\nu}}{M_H^2}\, T^V_{\mu\nu}\
 \frac{p_1^{\alpha} p_2^{\beta}}{M_H^2}\, T^V_{\alpha\beta}
\right]
\label{Xdef}
\eeqa
we can write
\beq
\sum_{pol} | {\cal M} |^2 = (gM_W)^2\, 
\frac{M_H^4}{4 M_W^4} \left[ \lambda(M_W^2,M_W^2;M_H^2) 
+12 \frac{M_W^4}{M_H^4} 
+ X(k_1,k_2,M_H,T^W) 
\right]
\label{m2hvv}
\eeq 

\noindent
It is easy to see that the 4--momenta $k_1$ and $k_2$ will only appear
in the square bracket of Eq.~(\ref{m2hvv}) as the scalar products like
$k_1 \cdot k_2$,  $P \cdot k_1$ and  $P \cdot k_2$.  These can all be
written in terms of the masses and therefore there is no angular
dependence in $d \Gamma$. Noticing also that 
\beq
|\vec k_1| = \frac{1}{2} M_H\, \sqrt{\lambda(M_W^2,M_W^2;M_H^2)}
\eeq
we can finally write
\beq
\Gamma=\frac{g^2 M_H^3}{64 \pi M_W^2} \sqrt{\lambda(M_W^2,M_W^2;M_H^2)}
\left[ \lambda(M_W^2,M_W^2;M_H^2) 
\! +\! \frac{M_W^4}{M_H^4} 
\! +\! X(k_1,k_2,M_H,T^W) 
\right]
\eeq
which can be written in terms of $G_F$ as
\beq
\Gamma=\frac{G_F M_H^3}{8 \pi \sqrt{2}}\, \sqrt{\lambda(M_W^2,M_W^2;M_H^2)}
\left[ \lambda(M_W^2,M_W^2;M_H^2) 
\!+\! 12 \frac{M_W^4}{M_H^4} 
\!+\! X(k_1,k_2,M_H,T^W) 
\right]
\eeq

\noindent
Now for the decay $H \ra Z^0 Z^0$ everything is similar except that we
have to divide by a factor of 2 because we have two identical
particles in the final state. Introducing $\delta_V=2(1)$ for
$V=W(Z)$, respectively, we can write both decays in a single formula

\beq
\Gamma=\delta_V
\frac{G_F M_H^3}{16 \pi \sqrt{2}}\, \sqrt{\lambda(M_V^2,M_V^2;M_H^2)}
\left[ \lambda(M_V^2,M_V^2;M_H^2) 
\!+\! 12 \frac{M_V^4}{M_H^4} 
\!+\!  X(k_1,k_2,M_H,T^V) 
\right]
\label{ghvv}
\eeq

\noindent
where $\lambda$ and $X$ are given in Eq.~(\ref{lambdadef}) and 
Eq.~(\ref{Xdef}). The SM part of Eq.~(\ref{ghvv}) agrees with Eq. (5)
of ref.~\cite{djouadi}. The term proportional to $X$ represents the
extra contributions from physics beyond the SM and is 
in agreement with the results of ref.~\cite{hagiwara} as we will
explain in Section~\ref{hagi}.

\subsection{The Off--Shell Decay $H\ra VV^*$}

We now consider the off--shell decay $H\ra VV^*$. To be precise we derive
the expression for $H\ra W^+ W^{-*} \ra W^+ f_i \ovl{f'}_i$ and 
then present a final result valid for all cases. 
We consider the kinematics given in Fig. 2.

\vbox{
\begin{picture}(0,5.5)
\put(4.5,0){\psfig{figure=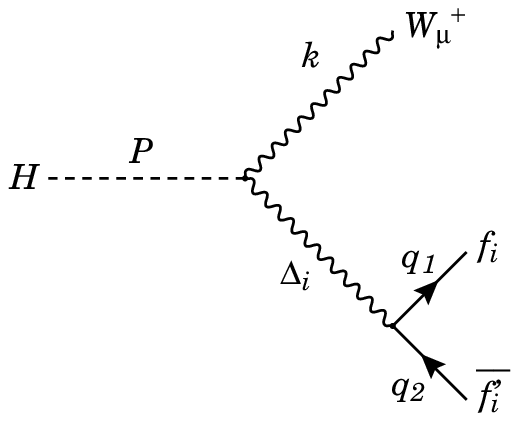}}
\end{picture}

\centerline{Fig. 2}
}

\noindent
where $(f_i,\ovl{f'}_i)$ represents one of the decay channels of the
$W^-$, for instance, $(e^-,\ovl{\nu}_e)$. Using the conventions of
ref.~\cite{PDG}, we can write the differential width as
\beq
d \Gamma= \frac{(2 \pi)^4}{2M_H} \sum_{pol} |{\cal M}|^2 d\Phi_3
\eeq
where $d\Phi_3$ is the phase space for 3 particles that we write
as~\cite{PDG}
\beq
d\Phi_3(P;k,q_1,q_2)=d\Phi_2(P;k,\Delta_i)\, d\Phi_2(\Delta_i;q_1,q_2)
(2\pi)^3 d \Delta_i^2
\eeq
with
\beq
\Delta_i=q_1+q_2 \quad ; \quad \Delta_i^2=(q_1 +q_2)^2
\eeq

\noindent
But the 2--body phase space in the rest frame of the decaying $W$ 
can be written as
\beq
d\Phi_2(\Delta_i;q_1,q_2)=
(2\pi)^{-6} \frac{|\vec {q^*}_1|}{4 \sqrt{\Delta_i^2}}\,  d \Omega^*_1 
=\frac{(2 \pi)^{-6}}{8}\, d \Omega^*_1
\eeq
where the last equality holds for massless decaying products of the
$W$ that we will assume 
and $\Omega^*_1$ is the solid angle of the particle with
momentum $q_1$ in the rest frame of the decaying $W$. Also the 2--body
phase space of the decaying $H$ can be written as
\beq
d \Phi_2(P;k,\Delta_i)
=\frac{(2 \pi)^{-6}}{8}\, \sqrt{\lambda(M_W^2,\Delta_i^2;M_H^2)}\, 
d \Omega_{\vec k}
\eeq
Putting everything together we get
\beq
d \Gamma =\frac{(2 \pi)^{-5}}{128 M_H}\, 
\sqrt{\lambda(M_W^2,\Delta_i^2;M_H^2)}\, 
\sum_{pol} | {\cal M}|^2\, 
d \Omega_{\vec k}\, d \Delta_i^2 d \Omega^*_1
\label{dghvv*}
\eeq

\noindent
Neglecting the fermion masses the matrix element ${\cal M}$ is 
\beq
{\cal M}= (gM_W) \epsilon^{\alpha}(k)\, \left(g_{\mu \alpha} 
+T^W_{\mu \alpha} \right) 
 \frac{1}{D(\Delta_i^2)}\,
\frac{g}{2 \sqrt{2}}\, \ovl{u}(q_1)\gamma^{\mu} \left(1
-\gamma_5\right) v(q_2)
\eeq
where
\beq
D(\Delta_i^2)= \Delta_i^2 -M_W^2 +i M_W \Gamma_W
\eeq

\noindent
We obtain for the matrix element squared
\beqa
\sum_{pol}|{\cal M}|^2&=& (gM_W)^2\, \frac{1}{|D(\Delta_i^2)|^2}\, 
\left(-g^{\alpha \beta} + \frac{k^{\alpha}k^{\beta}}{M_W^2}\right)
\left(g_{\mu \alpha} +T^W_{\mu \alpha} \right)
\left(g_{\nu \beta} +T^W_{\nu \beta} \right)\cr
\vb{24}
&&\hskip 3.5cm
\frac{g^2}{8}\, \hbox{Tr} \Big[\slash{q}_1 \gamma^{\mu} (1 -\gamma_5)
\slash{q}_2 \gamma^{\nu} (1 -\gamma_5) \Big]\cr
\vb{24}
&=&
(gM_W)^2\, \frac{1}{|D(\Delta_i^2)|^2}\, 
\left(-g^{\alpha \beta} + \frac{k^{\alpha}k^{\beta}}{M_W^2}\right)
\left(g_{\mu \alpha} +T^W_{\mu \alpha} \right)
\left(g_{\nu \beta} +T^W_{\nu \beta} \right)\cr
\vb{24}
&&\hskip 3.5cm
\frac{48 \pi\Gamma_i}{M_W} \, 
\Big[q_1^{\mu} q_2^{\nu} + q_2^{\mu} q_1^{\nu} 
-g^{\mu \nu} q_1 \cdot q_2 \Big] 
\label{m2hvv*}
\eeqa
where $\Gamma_i=g^2/(48 \pi)\, M_W$ is the decay width $W\ra f_i \ovl{f'}_i$.
Looking at Eq.~(\ref{dghvv*}) and Eq.~(\ref{m2hvv*}) we realize that
the only dependence on the solid angle $\Omega^*_1$ is inside the
square bracket in Eq.~(\ref{m2hvv*}). Then the integrals we have to
evaluate are of the form
\beq
I^{\alpha \beta} = \int d \Omega^*_1 q_1^{\alpha} q_2^{\beta}
\eeq
These can be easily done if we realize that in the rest frame of the
decaying $W$ the only 4--vector available is $\Delta_i$. We should
have then
\beq
I^{\alpha\beta}=
A \Delta_i^{\alpha} \Delta_i^{\beta} +B \Delta_i^2 g^{\alpha \beta}
\eeq
Multiplying the last equation respectively with $g_{\alpha \beta}$ and
with $\Delta_{i \alpha} \Delta_{i \beta}$ and noticing that 
$\Delta_i \cdot q_1=\Delta_i \cdot q_2= 1/2 \Delta_i^2$ we get a
system of equations for $A$ and $B$
\beq
\left\{\ba{l}
A +4 B = 2 \pi \cr
\vb{20}
A + B = \pi 
\ea
\right.
\eeq
which gives $A=2\pi/3$ and $B=\pi/3$. We get then
\beq
\int d \Omega^*_1 q_1^{\alpha} q_2^{\beta}=
\frac{\pi}{3} \left( 2 \Delta_i^{\alpha} \Delta_i^{\beta} 
+ \Delta_i^2 g^{\alpha \beta} \right)
\label{solidangle}
\eeq
and
\beq
\int d \Omega^*_1 \left[q_1^{\mu} q_2^{\nu} + q_2^{\mu} q_1^{\nu} 
-g^{\mu \nu} q_1 \cdot q_2 \right]=
\frac{4 \pi}{3} \left( \Delta_i^{\mu} \Delta_i^{\nu} 
- \Delta_i^2 g^{\mu \nu} \right)
\eeq

\noindent 
Doing the integration in $\Omega^*_i$ we get
\beqa
\int d \Omega^*_1\, \sum_{pol}|{\cal M}|^2&=& 
(gM_W)^2\, \frac{1}{|D(\Delta_i^2)|^2}\, 
\left(-g^{\alpha \beta} + \frac{k^{\alpha}k^{\beta}}{M_W^2}\right)
\left(g_{\mu \alpha} +T^W_{\mu \alpha} \right)
\left(g_{\nu \beta} +T^W_{\nu \beta} \right)\cr
\vb{24}
&&\hskip 3.5cm
\frac{48 \pi\Gamma_i}{M_W} \, 
\frac{4\pi}{3} \left( \Delta_i^{\mu} \Delta_i^{\nu} 
- \Delta_i^2 g^{\mu \nu} \right)
\label{m2hvv*2}
\eeqa

\noindent
If we compare Eq.~(\ref{m2hvv}) with Eq.~(\ref{m2hvv*2}) we can write
this last equation in the form

\beqa
\int d \Omega^*_1\, \sum_{pol} | {\cal M} |^2 &=& (gM_W)^2\,
 \frac{1}{|D(\Delta_i^2)|^2}\,  
\frac{M_H^4}{4 M_W^2}\, \frac{48 \pi\Gamma_i}{M_W} \, 
\frac{4\pi}{3} \cr 
\vb{24}
&&\hskip 0.5cm
\left[ \lambda(M_W^2,\Delta_i^2;M_H^2) 
+12 \frac{M_W^2 \Delta_i^2}{M_H^4} 
+  X(k,\Delta_i,M_H,T^W) 
\right]
\label{m2hvv*3}
\eeqa 

\noindent
We get therefore

\beqa
\frac{d \Gamma}{d \Delta_i^2 d \Omega_{\vec k}} 
&=&\frac{(2 \pi)^{-5}}{128 M_H}\, 
\sqrt{\lambda(M_W^2,\Delta_i^2;M_H^2)}\, 
(gM_W)^2\,
 \frac{1}{|D(\Delta_i^2)|^2}\,  
\frac{M_H^4}{4 M_W^2}\, \frac{48 \pi\Gamma_i}{M_W} \, 
\frac{4\pi}{3} \cr 
\vb{24}
&&\hskip 0.5cm
\left[ \lambda(M_W^2,\Delta_i^2;M_H^2) 
+12 \frac{M_W^2 \Delta_i^2}{M_H^4} 
+  X(k,\Delta_i,M_H,T^W) 
\right]
\label{dghvv*2}
\eeqa

\noindent
Next we realize that in Eq.~(\ref{dghvv*2}) there is no dependence on
the solid angle of the real $W$. We can therefore trivially perform
that integration. We get
\beqa
\frac{d \Gamma}{d \Delta_i^2} 
&=&\frac{(2 \pi)^{-5}}{128 M_H}\, (4 \pi)\, 
\sqrt{\lambda(M_W^2,\Delta_i^2;M_H^2)}\, 
(gM_W)^2\,
 \frac{1}{|D(\Delta_i^2)|^2}\,  
\frac{M_H^4}{4 M_W^2}\, \frac{48 \pi\Gamma_i}{M_W} \, 
\frac{4\pi}{3} \cr 
\vb{24}
&&\hskip 0.5cm
\left[ \lambda(M_W^2,\Delta_i^2;M_H^2) 
+12 \frac{M_W^2 \Delta_i^2}{M_H^4} 
+  X(k,\Delta_i,M_H,T^W) 
\right]
\label{dghvv*3}
\eeqa
and finally we get for the width
\beqa
\Gamma
&=&\frac{G_F M_H^3}{8 \pi \sqrt{2}}\,
\sqrt{\lambda(M_W^2,\Delta_i^2;M_H^2)}\, 
 \frac{1}{\pi} \int d \Delta_i^2 
\frac{\Gamma_i M_W}{|D(\Delta_i^2)|^2}\, \cr  
\vb{24}
&&\hskip 0.5cm
\left[ \lambda(M_W^2,\Delta_i^2;M_H^2) 
+12 \frac{M_W^2 \Delta_i^2}{M_H^4} 
+  X(k,\Delta_i,M_H,T^W) 
\right]
\label{ghvv*}
\eeqa

\noindent
or

\beq
\Gamma=
\frac{1}{\pi} \int d \Delta_i^2 
\frac{\Gamma_i M_W}{|D(\Delta_i^2)|^2}\, \Gamma_0^W(k,\Delta_i,M_H)
\label{ghvv*2}
\eeq

\noindent
where

\beqa
\Gamma_0^W(k,\Delta_i,M_H)&=&
\frac{G_F M_H^3}{8 \pi \sqrt{2}}\,
\sqrt{\lambda(M_W^2,\Delta_i^2;M_H^2)}\,  
\left[\vb{18} \lambda(M_W^2,\Delta_i^2;M_H^2) \right.\cr
\vb{24}
&&\left. \hskip 1cm
+12 \frac{M_W^2 \Delta_i^2}{M_H^4} 
+  X(k,\Delta_i,M_H,T^W) 
\right]
\eeqa

\noindent
If we sum over all the final states of the $W$ we can substitute
$\Gamma_i$ with $\Gamma_W$. Eq.~(\ref{ghvv*2}) is in agreement with
Eq.~(6) of ref.~\cite{marciano} in the zero width limit.
Similar considerations apply to the case
of the decay $H \ra Z^0 + f_i \ovl{f}_i$. We can summarize the final
result in the formula,

\beq
\Gamma=
\frac{1}{\pi} \int d \Delta_i^2 
\frac{\Gamma_V M_V}{|D(\Delta_i^2)|^2}\, \Gamma_0^V(k,\Delta_i,M_H)
\label{ghvv*3}
\eeq

\noindent
where

\beqa
\Gamma_0^V(k,\Delta_i,M_H)&=& \delta_V
\frac{G_F M_H^3}{16 \pi \sqrt{2}}\,
\sqrt{\lambda(M_V^2,\Delta_i^2;M_H^2)}\,  
\left[\vb{18} \lambda(M_V^2,\Delta_i^2;M_H^2) \right.\cr
\vb{24}
&&\left. \hskip 1cm
+12 \frac{M_V^2 \Delta_i^2}{M_H^4} 
+  X(k,\Delta_i,M_H,T^V) 
\right]
\eeqa

\noindent
$\delta_V$ was defined before, $X$ is given in Eq.~(\ref{Xdef})
 and $k^2=M_V^2$.

\subsection{The Off--Shell Decay $H\ra V^*V^*$}

We now consider the off--shell decay $H\ra V^*V^*$. To be precise we derive
the expression for 
$H\ra W^{+*} W^{-*} \ra (f_i \ovl{f'}_i) + (f_j \ovl{f'}_j)$ and 
then present a final result valid for all cases. 
We consider the kinematics given in Fig. 3.

\vbox{
\begin{picture}(0,5.5)
\put(4.5,0){\psfig{figure=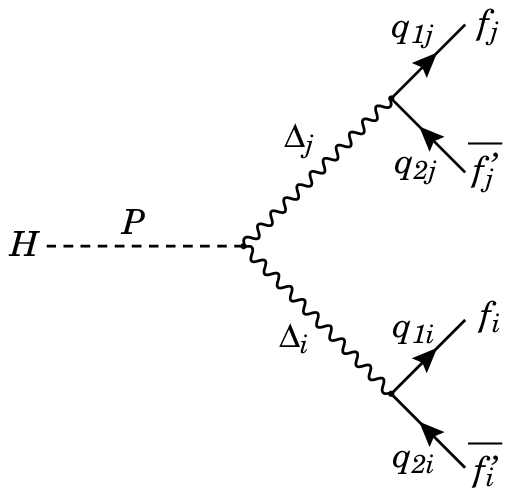}}
\end{picture}

\centerline{Fig. 3}
}

\noindent
where $(f_i,\ovl{f'}_i)$ represents one of the decay channels of the
$W^-$ and $(f_j,\ovl{f'}_j)$ represents one of the decay channels of the
$W^+$. After we have done the case $H\ra V V^*$ it is very easy to do
this case. 

\noindent
The expression for the width is~\cite{PDG},
\beq
d \Gamma= \frac{(2 \pi)^4}{2M_H} \sum_{pol} |{\cal M}|^2 d\Phi_4
\eeq
where $d\Phi_4$ is the phase space for 4 particles that we write
as~\cite{PDG}
\beq
d\Phi_4(P;k,q_1,q_2)=d\Phi_2(P;\Delta_i,\Delta_j)\, 
d\Phi_2(\Delta_i;q_{i1},q_{i2}) (2\pi)^3 d \Delta_i^2\,
d\Phi_2(\Delta_j;q_{j1},q_{j2}) (2\pi)^3 d \Delta_j^2
\eeq
with
\beq
\Delta_i=q_{i1}+q_{i2} \ ; \ \Delta_i^2=(q_{i1} +q_{i2})^2
\quad ; \quad
\Delta_j=q_{j1}+q_{j2} \ ; \ \Delta_j^2=(q_{j1} +q_{j2})^2
\eeq

\noindent
But the 2--body phase spaces can be written as
\beqa
d\Phi_2(\Delta_i;q_{i1},q_{i2})
&=&\frac{(2 \pi)^{-6}}{8}\, d \Omega^*_{i1} \cr
\vb{24}
d\Phi_2(\Delta_j;q_{j1},q_{j2})
&=&\frac{(2 \pi)^{-6}}{8}\, d \Omega^*_{j1} \cr
\vb{24}
d \Phi_2(P;\Delta_i,\Delta_j)
&=&\frac{(2 \pi)^{-6}}{8}\, \sqrt{\lambda(\Delta_i^2,\Delta_j^2;M_H^2)}\, 
d \Omega_{\vec \Delta_i}
\eeqa
where, as before, we consider that the decays products of the
$W^{\pm}$ are massless. Putting everything together we  have

\beq
d \Gamma= \frac{(2\pi)^{-8}}{2^{10}M_H}\, 
\sqrt{\lambda(\Delta_i^2,\Delta_j^2;M_H^2)}\, 
\sum_{pol}|{\cal M}|^2 d \Delta_i^2\, d \Delta_j^2\,
d \Omega_{\vec \Delta_i}\,  d \Omega^*_{i1}  d \Omega^*_{j1} 
\eeq

\noindent
The matrix element is 

\beqa
{\cal M}&=& 
(gM_W)\, \frac{1}{D(\Delta_i^2)}\,
\frac{1}{D(\Delta_j^2)}\,
\frac{g}{2\sqrt{2}}\ \ovl{u}(q_{i1}) \gamma^{\mu}(1 -\gamma_5)
v(q_{i2})\cr
\vb{22}
&&
\frac{g}{2\sqrt{2}}\ \ovl{u}(q_{j1}) \gamma^{\mu}(1 -\gamma_5)
v(q_{j2})
\eeqa
and the same procedure that we used for the $H\ra VV^*$ case gives

\beqa
\int d \Omega^*_{1i}\, d\Omega^*_{1j}
\sum_{pol} | {\cal M} |^2 &=& (gM_W)^2\,
\frac{1}{|D(\Delta_i^2)|^2}\,  \frac{1}{|D(\Delta_j^2)|^2}\,  
\frac{M_H^4}{4 }\, \frac{(48 \pi)^2\Gamma_i \Gamma_j}{M_W^2}\, 
\left(\frac{4\pi}{3}\right)^2 \cr 
\vb{24}
&&
\left[ \lambda(\Delta_i^2,\Delta_j^2;M_H^2) 
+12 \frac{\Delta_i^2 \Delta_j^2}{M_H^4} 
+  X(\Delta_i,\Delta_j,M_H,T^W) 
\right]
\label{m2hv*v*}
\eeqa 

\noindent
and after doing the $d \Omega_{\vec \Delta_i}$ integration we obtain

\beqa
\Gamma
&=&\frac{G_F M_H^3}{8 \pi \sqrt{2}}\,
\sqrt{\lambda(\Delta_i^2,\Delta_j^2;M_H^2)}\, 
 \frac{1}{\pi} \int d \Delta_i^2 
\frac{\Gamma_i M_W}{|D(\Delta_i^2)|^2}\, 
 \frac{1}{\pi} \int d \Delta_j^2 
\frac{\Gamma_j M_W}{|D(\Delta_j^2)|^2}\, \cr  
\vb{24}
&&\hskip 0.5cm
\left[ \lambda(\Delta_i^2,\Delta_j^2;M_H^2) 
+12 \frac{\Delta_i^2 \Delta_i^2}{M_H^4} 
+  X(\Delta_i,\Delta_j,M_H,T^W) 
\right]
\label{ghv*v*}
\eeqa

\noindent
Summing over all final states we get

\beq
\Gamma=
\frac{1}{\pi} \int d \Delta_i^2 \frac{\Gamma_V
M_V}{|D(\Delta_i^2)|^2}\, \frac{1}{\pi}
\int d \Delta_j^2 \frac{\Gamma_V
M_V}{|D(\Delta_j^2)|^2}\, 
\Gamma_0^V(\Delta_i,\Delta_j,M_H)
\label{ghv*v*2}
\eeq

\noindent
where\footnote{One might worry about the factor $\delta_V$. For the
$W^{+*}W^{-*}$ final case there is no problem because the final states
of the $W^{+*}$ are different from the final states of the
$W^{-*}$. Therefore
\beqa
\sum_{Final\ States} \Gamma
\left[H\ra (W^{+*}\ra i_+)+(W^{-*}\ra i_-) \right]
\vb{20}
&\propto& \sum_{i_+} \sum_{i_-} \Gamma(W^{+*}\ra i_+)
\Gamma(W^{-*}\ra i_-) \cr
\vb{20}
&=&\sum_{i_+} \Gamma(W^{+*}\ra i_+)\
 \sum_{i_-} \Gamma(W^{-*}\ra i_-)\cr
\vb{20}
&=&\Gamma_W \Gamma_W
\eeqa
and $\delta_W=2$ because of the constants we factored out. 
For the $H\ra (Z^*\ra i)+(Z^*\ra j)$ case one should
be more careful. If $i\not=j$ than we should divide by 2 otherwise we
would be double counting in the product 
$(\Gamma_1 +\Gamma_2 +\cdots)(\Gamma_1 +\Gamma_2 +\cdots)$. For $i=j$
there is no double counting in the above product, but now we have two
pairs of identical particles in the final state but we also have 2
diagrams. Then we should square the sum of the amplitudes and divide
by 4. In general this would not give a factor of 1/2 because of the
interference term. However the interference will be negligible because
the momenta squared in the denominators cannot be equal to $M_Z$ in
all 4 lines (of the product of the 2 diagrams) at the same
time. Therefore if we neglect the interference we should divide also
by 2 in this case. Therefore $\delta_Z=1$.
}

\beqa
\Gamma_0^V(\Delta_i,\Delta_j,M_H)&=& \delta_V
\frac{G_F M_H^3}{16 \pi \sqrt{2}}\,
\sqrt{\lambda(\Delta_i^2,\Delta_j^2;M_H^2)}\,  
\left[\vb{18} \lambda(\Delta_i^2,\Delta_j^2;M_H^2) \right.\cr
\vb{24}
&&\left. \hskip 1cm
+12 \frac{\Delta_i^2 \Delta_j^2}{M_H^4} 
+  X(\Delta_i,\Delta_j,M_H,T^V) 
\right]
\eeqa

\ni
This result is in agreement with \cite{ee500,djouadi}, except for the
value of $\delta_Z$. One should mention that formulas for off--shell
decays of the type of Eqs.~(\ref{ghvv*3}) and (\ref{ghv*v*2}) for
other decays are known in the literature~\cite{off-shell}.

\subsection{A comparison of the various formulas}

Perhaps it is useful to indicate the domain of validity of the various
formulas for the widths. This will depend on the value of the Higgs
boson mass. In Figure~\ref{whww} we plot the various formulas for the case of
$H\ra W^+W^-$.
\setcounter{figure}{3}
\begin{figure}
\begin{picture}(0,10)
\put(3.5,0){\psfig{figure=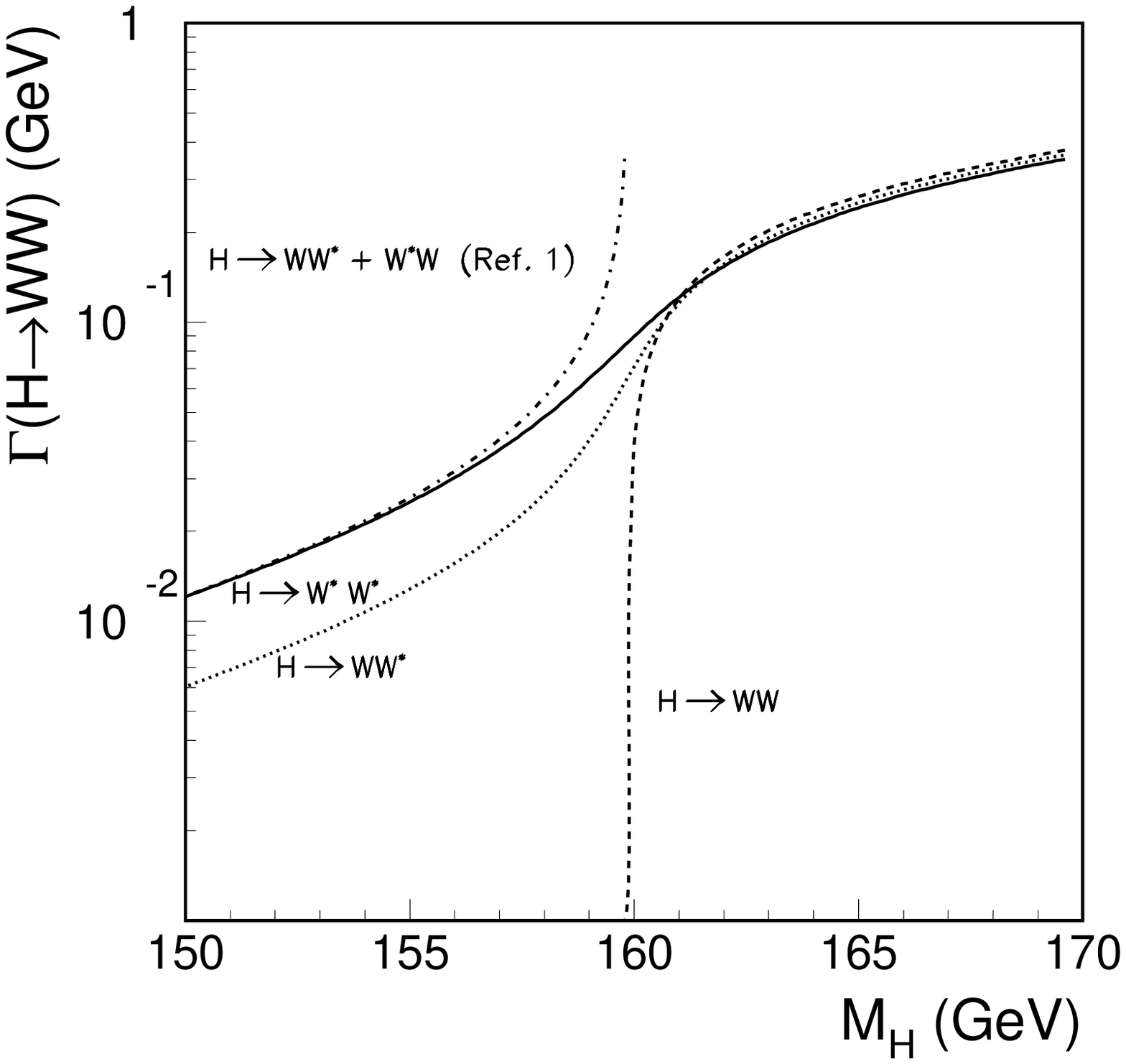,height=9cm}}
\end{picture}
\caption{\small
Comparison of the off--shell and on--shell formulas for $H \ra W^+W^-$.
The dashed line corresponds to the on--shell formula
Eq.~(\ref{ghvv}), the dotted line to the case that only one W is off-shell
Eq.~(\ref{ghvv*}), and the solid line corresponds to the case where both W's
are off--shell Eq.~(\ref{ghv*v*}). For comparison is 
also shown Eq.~(6) of ref.~\cite{marciano}. 
}
\label{whww}
\end{figure}
From this figure it is clear that the proper way to calculate the
width below the two $W$'s threshold is to use Eq.~(\ref{ghv*v*}) with
the two $W$'s off--shell. The two integrations in Eq.~(\ref{ghv*v*})
automatically take care of the fact that either one of the $W$'s can be
close to be on--shell. In Eq.~(6) of ref.~\cite{marciano} this is done by
adding the two possibilities, but as the width is neglected the formula
is only good below the threshold.

\section{The Decay $H\ra \gamma Z$}
\label{SecHGZ}

Due to the electromagnetic gauge invariance the most general
expression for the coupling $ H \gamma Z$ is

\vskip 1.5cm
\beq
\hskip 5cm -i\, \frac{e^2g}{16 \pi^2 M_W}\ (g_{\mu \nu}\, k \cdot q - k_{\nu}
q_{\mu} )\, A(q^2,M_H)
\label{haz_coupling}
\eeq

\begin{picture}(0,0.2)
\put(1.5,-1){\psfig{figure=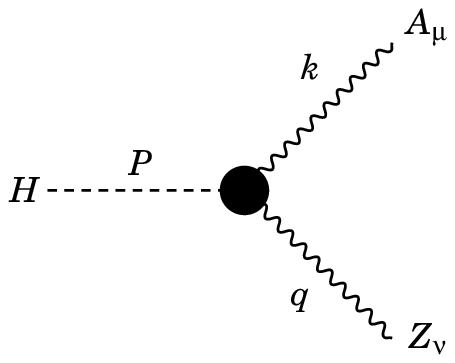}}
\end{picture}
\bs

\centerline{Fig. 5}

\ni
where $A(q^2,M_H)$ is a dimensionless form factor that depends only in
the mass of the $H$ and on the square of the momentum of the $Z$ (if
the $Z$ is on--shell then $q^2=M_Z^2$). In the SM the lowest
contribution to $A$ is at the 1--loop level. If we are considering
physics Beyond the Standard Model (BSM) then we should have
\beq
A=A_{SM} + A_{BSM}
\eeq
where the SM contribution is given~\cite{HHG,barroso,babu} by
\beq
A_{SM}= A_W + A_F
\label{A_SM}
\eeq
with\footnote{Our convention here for the coupling,
Eq.~(\ref{haz_coupling}) is as in ref.~\cite{HHG}. It differs from our
previous convention, ref.~\cite{babu},  by a factor of $-1/\sin
\theta_W$. Our conventions are explained in Appendix A.} 
\beqa
A_W&=&-4 \cot \theta_W \left[\vbox to 18pt{}
(3 -\tan \theta_W^2)\ J_1(q^2,M_H^2,M_W^2) \right. \cr
&& \cr
&& \left. + \left(-5 +\tan_W^2 \theta_W - \frac{1}{2} 
\frac{M_H^2}{M_W^2}(1-\tan \theta_W^2) \right)\ J_2(q^2,M_H^2,M_W^2)
\right]
\eeqa
and
\beq
A_F=-\sum_f \frac{4 g^f_V Q_f}{\sin \theta_W \cos \theta_W}\ 
\left[\vbox to 14pt{}
-J_1(q^2,M_H^2,M_f^2)+4 J_2(q^2,M_H^2,M_f^2) \right] \ .
\eeq
where $Q_f$ is the charge, in units of $|e|$, of the fermion $f$ in
the loop, and
$g^f_V=1/2 T_3^f -Q_f \sin^2 \theta_W$. 
The explicit form of the functions $J_1$ and $J_2$  can be found in
Appendix B.
In the following we will use this general coupling to evaluate both
the on--shell and the off--shell decays of the Higgs boson.

\subsection{The On--Shell Decay $H\ra \gamma Z$}

The differential width is, like before (see Eq.~(\ref{dghvv}))

\beq
d \Gamma= \frac{1}{32 \pi^2}\, \sum_{pol} | {\cal M}|^2\, \frac{|\vec k
|}{M_H^2}\, d \Omega_{\vec k}
\label{dghaz}
\eeq
where 
\beq
{\cal M}= \epsilon^{\mu}(k) \epsilon^{\nu}(q)\,
\frac{e^2 g}{16 \pi^2 M_W}\ (g_{\mu \nu}\, k \cdot q - k_{\nu}
q_{\mu} )\, A(q^2,M_H)
\label{mhaz}
\eeq
We get therefore
\beq
\sum_{pol} | {\cal M} |^2 = \left( \frac{e^2 g}{16 \pi^2 M_W}\right)^2\, 
2 (k \cdot q)^2\, |A|^2
\label{m2haz}
\eeq
Now using 
\beq
|\vec k|= \frac{k \cdot q}{M_H}= \frac{1}{2}\, M_H \sqrt{\lambda(M_Z^2,0;M_H^2)}
\eeq
where
\beq
\lambda(M_Z^2,0;M_H^2)=\left(1 -\frac{M_Z^2}{M_H^2} \right)^2
\eeq
we get finally
\beq
\Gamma=
\frac{G_F M_H^3}{4 \pi \sqrt{2}}\, 
\frac{\alpha^2}{16 \pi^2 }\, 
\lambda(M_Z^2,0;M_H^2)^{3/2}\, |A|^2
\label{ghav}
\eeq
\ni
This result is in agreement for the SM with refs.~\cite{HHG,barroso}
but it differs by a factor of two from ref.~\cite{hagiwara} that
claims to have the same definition of $A$ as we and ref.~\cite{HHG} do. 

\subsection{The Off--Shell Decay $H\ra \gamma Z^*$}

We consider for definiteness the the decay $H\ra \gamma f_i \ovl{f}_i$
as represented in Fig. 6

\vbox{
\begin{picture}(0,5)
\put(5,0){\psfig{figure=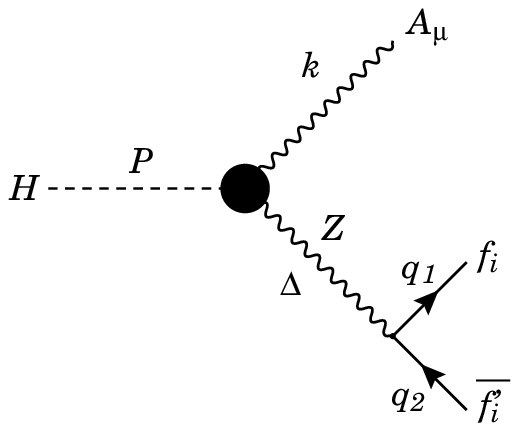}}
\end{picture}

\centerline{Fig. 6}
}

\ni
The differential width can be written as in Eq.~(\ref{dghvv*})

\beq
d \Gamma =\frac{(2 \pi)^{-5}}{128 M_H}\, 
\sqrt{\lambda(0,\Delta^2;M_H^2)}\, 
\sum_{pol} | {\cal M}|^2\, 
d \Omega_{\vec k}\, d \Delta^2 d \Omega^*_1
\label{dghav*}
\eeq

\noindent
where the matrix element ${\cal M}$ is (we again neglect the fermion
masses)   
\beq
{\cal M}= \epsilon^{\mu}(k) 
\frac{e^2g}{16 \pi^2 M_W}\ (g_{\mu \nu}\, k \cdot \Delta - k_{\nu}
\Delta_{\mu} )\, A(\Delta^2,M_H)\,
\frac{1}{D(\Delta^2)}\,
\frac{g}{\cos \theta_W} \ovl{u}(q_1)\gamma^{\nu} (g_V^f -g_A^f
\gamma_5) v(q_2)
\eeq
and
\beq
\Delta=q_1 +q_2
\eeq
Our conventions for the couplings of the $Z$ to the fermion $f$ are
given in Appendix A.
The sum over polarizations and spins of the matrix element squared
gives now
\beqa
\sum_{pol} |{\cal M}|^2 &=& \left(\frac{e^2g}{16\pi^2 M_W)}\right)^2\,
\frac{1}{|D(\Delta^2)|^2}\,
\left(\frac{g}{\cos \theta}\right)^2\,
8 |A|^2\, \left(g_V^f{}^2+g_A^f{}^2 \right)\cr
\vb{22}
&& \left[\vb{16}  k \cdot \Delta\ k \cdot q_1\ \Delta \cdot q_2 
         + k \cdot \Delta\ k \cdot q_2\ \Delta \cdot q_1 
         - k \cdot q_1\ k \cdot q_2\ \Delta \cdot \Delta \right]
\eeqa
Using now Eq.~(\ref{solidangle}) to perform the integration over 
the solid angle in the center of mass frame of the decaying $Z$ we get
\beq
\int d\Omega^*_1 \sum_{pol} |{\cal M}|^2=
\left(\frac{e^2g}{16\pi^2 M_W}\right)^2\,
\frac{1}{|D(\Delta^2)|^2}\,
\left(\frac{g}{\cos \theta}\right)^2\,
|A|^2\, \left(g_V^f{}^2+g_A^f{}^2 \right)\, 
\frac{32\pi}{3}\, (k\cdot \Delta)^2\, \Delta^2
\eeq
We can now perform the integration over the solid angle of the photon
and obtain
\beq
\frac{d\Gamma}{d\Delta^2}=
\frac{1}{32\pi^2M_H}\, 
\lambda(\Delta^2,0;M_H^2)^{3/2}\, M_H^3\,
\left(\frac{eg^2}{16\pi^2 M_W}\right)^2\,
\frac{\Gamma_i}{M_Z}\, \Delta^2 \,
\frac{1}{|D(\Delta^2)|^2}
\eeq
where we have used the expression for the partial width $\Gamma_i$ of
$Z\ra f_i \ovl{f}_i$ 
\beq
\Gamma_i=\frac{1}{12\pi}\,
\left(\frac{g}{\cos \theta_W}\right)^2 \,
\left(g_V^f{}^2+g_A^f{}^2 \right)
\eeq
Summing over all the final states we obtain finally
\beq
\Gamma=
\frac{1}{\pi}\, \int d\Delta^2\
\frac{\Gamma_Z}{M_Z} \,
\frac{ \Delta^2}{|D(\Delta^2)|^2}\
\Gamma^{\gamma Z}(M_H,\Delta^2)
\label{ghav*}
\eeq
where
\beq
\Gamma^{\gamma Z}(M_H,\Delta^2)=
\frac{G_F M_H^3}{4 \pi \sqrt{2}}\, 
\frac{\alpha^2}{16 \pi^2 }\, 
\lambda(\Delta^2,0;M_H^2)^{3/2}\, |A(\Delta^2,M_H^2)|^2
\eeq
is the decay into an off-shell $Z$ and gives back Eq.~(\ref{ghav})
when $\Delta^2=M_Z^2$.

\section{The decay $H\ra \gamma \gamma$}

For completeness we also give the general formula for this decay. 
Due to the electromagnetic gauge invariance the most general
expression for the coupling\footnote{We are assuming that only one photon
is off-shell because this is case of interest.} $ H \gamma^* \gamma$ is

\vskip 1.5cm
\beq
\hskip 5cm i\, \frac{e^2g}{16 \pi^2 M_W}\ (g_{\mu \nu}\, k \cdot q - k_{\nu}
q_{\mu} )\, I(q^2,M_H)
\label{haa_coupling}
\eeq

\vbox{
\begin{picture}(0,0.2)
\put(1.5,-1){\psfig{figure=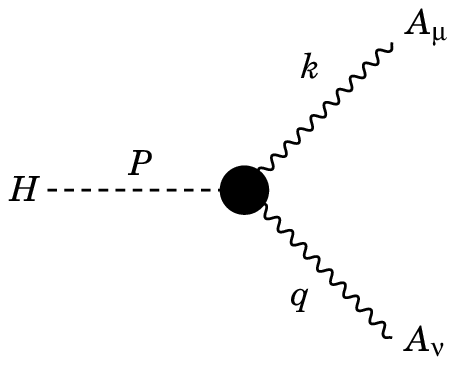}}
\end{picture}
\bs

\centerline{Fig. 7}
}

\ni
where, as before,
\beq
I=I_{SM} + I_{BSM}
\eeq
The Standard Model contribution is given by~\cite{hagiwara,HHG,barroso}
\beq
I_{SM}= I_W + I_F
\label{I_SM}
\eeq
where
\beq
I_W=-4 \left[\vbox to 18pt{}
-4 J_1(q^2,M_H^2,M_W^2) 
+ \left(6 + \frac{M_H^2}{M_W^2}\right)\ J_2(q^2,M_H^2,M_W^2)
\right]
\eeq
and
\beq
I_F=\sum_f 4 Q^2_f\ 
\left[\vbox to 14pt{}
-J_1(q^2,M_H^2,M_f^2)+4 J_2(q^2,M_H^2,M_f^2) \right] \ .
\eeq
Using the above coupling and comparing with the case $H \ra \gamma Z$,
Eqs.~(\ref{dghaz}),~(\ref{mhaz}) e (\ref{m2haz}), 
it is straightforward to obtain
\beq
\Gamma(H\ra \gamma \gamma)= 
\frac{\alpha^3 M_H^3}{256 \pi^2 \sin^2 \theta_W M_W^2}\
|I|^2
\eeq
in agreement with refs.~\cite{HHG,hagiwara}.

\section{The  3--point functions $Z^*\ra H \gamma$ and
$\gamma^* \ra H \gamma$}

For some applications it is also important to know the related off--shell
3-point functions $Z^*\ra H \gamma$ and $\gamma^* \ra H \gamma$. For
completeness we collect them here.

\subsection{The $Z^*\ra H \gamma$ 3--point function}

We use the results of refs.~\cite{barroso,babu}. The amplitude can be
written as\footnote{Notice that our conventions here differ by a 
factor $-1/\sin\theta_W$ with respect to ref.~\cite{babu}.}
\beq
i\, {\cal M}=i\, \epsilon_Z^{\nu}(q)\,
\epsilon_A^{\mu}(k)\,
\left(\frac{e^2 g}{16\pi^2M_W}\right)\,
\left(g_{\mu \nu}\, k \cdot q - k_{\nu} q_{\mu} \right)\ A(q^2,M_H)
\label{zha_coupling}
\eeq
where
\beq
A = A_{SM} + A_{BSM}
\eeq
The standard model dimensionless amplitude $A_{SM}$ is given by
Eq.~(\ref{A_SM}). The sign difference between Eq.~(\ref{A_SM}) and 
Eq.~(\ref{zha_coupling}) is due to the fact that in the first one $q$ is an 
{\it outgoing} momentum and in the last one  is {\it incoming}.

\subsection{The $\gamma^*\ra H \gamma$ 3--point function}

Again using ref.~\cite{barroso} we have
\beq
i\, {\cal M}=-i\, \epsilon_A^{\nu}(q)\,
\epsilon_A^{\mu}(k)\,
\left(\frac{e^2 g}{16\pi^2M_W}\right)\,
\left(g_{\mu \nu}\, k \cdot q - k_{\nu} q_{\mu} \right)\ I(q^2,M_H)
\label{aha_coupling}
\eeq
where
\beq
I = I_{SM} + I_{BSM}
\eeq
The standard model dimensionless amplitude $I_{SM}$ is given by
Eq.~(\ref{I_SM}).

\subsection{An Effective Lagrangian for the SM Couplings}

We can write an effective Lagrangian that reproduces the couplings
given in Eqs.~(\ref{haz_coupling}),~(\ref{haa_coupling}),
~(\ref{zha_coupling}), ~(\ref{aha_coupling}). This is specially useful
if we want to add new physics, in addition to the SM, as we will show
in the next section. We get

\beq 
{\cal L}^{eff}=-\frac{1}{4}\, A_{\mu \nu} A^{\mu \nu} H\ {\cal I}_{SM}
+\frac{1}{2}\, A_{\mu \nu} Z^{\mu \nu} H\ {\cal A}_{SM}
\label{lageff}
\eeq

\ni
where we have defined
\beq
A_{\mu \nu}=\partial_{\mu} A_{\nu} -\partial_{\nu} A_{\mu}
\quad ; \quad
Z_{\mu \nu}=\partial_{\mu} Z_{\nu} -\partial_{\nu} Z_{\mu}
\eeq
and
\beqa
{\cal I}_{SM}&=& \frac{e^2 g}{16 \pi^2 M_W}\ I_{SM}(q^2,M_H) \cr
\vb{22}
{\cal A}_{SM}&=&\frac{e^2 g}{16 \pi^2 M_W}\ A_{SM}(q^2,M_H)
\eeqa
The effective Lagrangian, Eq.~(\ref{lageff}), is valid for the case of
one on--shell photon, the other photon (or $Z^0$) can be either
on--shell or off-shell. $A_{SM}$ and $I_{SM}$ are given in
Eqs.~(\ref{A_SM}) and (\ref{I_SM}).

\section{An example of extension of the SM}
\label{hagi}

A possible enhancement of the production and decay rates of the Higgs
boson can be originated by an  anomalous couplings of the Higgs boson to
the vector bosons. These interactions can be described in terms
of an effective dimension-six term in the interaction Lagrangian density

\begin{equation}
{\cal L}_{eff} = \sum_{i=1}^{7}   \frac{f_i}{\Lambda^2} O_i
\label{eq:lagrange}
\end{equation}
\noindent 
where the $O_i$ are the operators which represent the anomalous couplings,
$\Lambda$ is the typical energy scale of the interaction and $f_i$
are the constants which define the strength of each term 
\cite{hagiwara2,hagiwara}.

\ni
The anomalous couplings $H\gamma\gamma$, $HZZ$, $HZ\gamma$ and $HWW$
follow from the effective Lagrangian (\ref{eq:lagrange})
and can be written in the unitary gauge \cite{hagiwara2,hagiwara} as,
\begin{eqnarray}
{\cal L}_{eff}^{HVV} & = & g\frac{m_W}{\Lambda^2}   \left[
- \frac{s^2(f_{BB} + f_{WW} - f_{BW})}{2}HA_{\mu\nu}A^{\mu\nu}
+ \frac{2m^2_W}{g^2}\frac{f_{\phi ,1}}{c^2}HZ_{\mu}Z^{\mu} \right.
\nonumber \\
\vb{24}
& + & \frac{c^2f_W+s^2f_B}{2c^2}Z_{\mu\nu}Z^{\mu}(\partial^\nu H)
-\frac{s^4f_{BB}+c^4f_{WW}+s^2c^2f_{BW}}{2c^2}HZ_{\mu\nu}Z^{\mu\nu}
\nonumber \\
\vb{24}
& + & \frac{s(f_W-f_B)}{2c}A_{\mu\nu}Z^{\mu}(\partial^\nu H)
+\frac{s(2s^2f_{BB}-2c^2f_{WW}+(c^2-s^2)f_{BW})}{2c}HA_{\mu\nu}Z^{\mu\nu}
\nonumber \\
\vb{24}
& +&\left.
\frac{f_W}{2}(W^+_{\mu\nu}W^{-\mu}+W^-_{\mu\nu}W^{+\mu})(\partial^{\nu}H)
-f_{WW}HW^+_{\mu\nu}W^{-\mu\nu}\right]
\end{eqnarray}
where $X_{\mu\nu} = \partial_\mu X_\nu - \partial_\nu X_\mu$ with
$X=A,Z,W$, and $s(c)=\sin \theta_W (\cos \theta_W)$, respectively.

\ni
Both $f_{\phi ,1 }$ and $f_{BW}$ are already severely constrained by
precise measurements at low energy experiments, once they contribute to
the $Z^0$ mass and to the $B-W^3$ mixing, respectively. In what follows
these parameters will be assumed to be zero. 
Under this assumption, both HWW and HZZ have the same tensorial
structure. With the convention $H(p_H) \ra V^{\mu}(p_1) +
V^{\nu}(p_2)$ for the momenta, we have:
\beqa
T_V^{\mu \nu} &\equiv&
- A_V \left[ \vb{12} p_1^{\nu} p_H^{\mu}-(p_1 . p_H)g^{\nu \mu}+
p_H^{\nu} p_2^{\mu}-(p_2 . p_H)g^{\nu \mu}\right]+
B_V \left[ \vb{12} p_1^{\nu} p_2^{\mu} -(p_1 . p_2)g^{\nu \mu} \right]
\eeqa
(V = Z, W), where :
\beqa
A_W &\equiv& - \frac{1}{2} \frac{f_W}{\Lambda^2} \cr
\vb{20}
B_W &\equiv& - 2  \frac{f_{WW}}{\Lambda^2} \cr
\vb{20}
A_Z &\equiv& - \frac{1} {2} (\frac{f_B}{\Lambda^2}
\sin^2 \theta_W + \frac{f_W}{\Lambda^2} \cos^2 \theta_W)\cr
\vb{20}
B_Z &\equiv& - 2 (\frac{f_{BB}}{\Lambda^2}
\sin^4 \theta_W + \frac{f_{WW}}{\Lambda^2} \cos^4 \theta_W)
\label{AVBV}
\eeqa

\ni
The value of $X_V(p_1,p_2,M_H,T^V)$ as defined in Eq.~(\ref{Xdef})
is, thus, given by:
\beqa
X_V  & = &
4 \left\{ A_V \;\ \left[ \;\ 4 \frac{p_1^2 p_2^2}{M_H^2}
- \frac{p_1 . p_2}{M_H^4}( \,\ (p_1^2 - p_2^2)^2 - ( p_1^2 + p_2^2 ) M_H^2
\,\ ) \;\ \right]  \right. \cr
\vb{24}
&&\hskip 0.5cm
+ B_V \;\ \left[ \;\ -6 \frac{ \,\ (p_1 .  p_2) \,\ p_1^2 p_2^2}{M_H^4}
\;\ \right] 
\cr
\vb{24}
&&\hskip 0.5cm
+ A^2_V \;\ \left[\;\ p_1^2 p_2^2 +
\frac{(p_1^2+p_2^2)( 4  p_1^2 p_2^2 - (M_H^2 -
(p_1^2 + p_2^2) \,\ )^2)}{4  M_H^2} \right.
\cr
\vb{24}
&&\hskip 1.5cm\left.
+ \frac {(M_H^4 - ( p_1^2 - p_2^2 )^2) (4 p_1^2 p_2^2 + M_H^2 (p_1^2 +
p_2^2) - (p_1^2 + p_2^2)^2)}{4  M_H^4} \;\ \right] 
\cr
\vb{24}
&&\hskip 0.5cm
+
A_V B_V \;\ \left[ \;\ -2 \frac{p_1^2 p_2^2  (M_H^2 -(p_1^2 + p_2^2))}{M_H^2} +
\frac { p_1^2 p_2^2 ( (p_1^2 - p_2^2)^2 - M_H^2 (p_1^2 + p_2^2)
)}{M_H^4} \;\ \right] 
\cr
\vb{24}
&&\hskip 0.5cm \left.
+ B^2_V \;\ \left[ \;\ \frac{p_1^2 p_2^2}{2 M_H^4} ( ( M_H^2 -
(p_1^2 + p_2^2) \,\ )^2 + 2 p_1^2 p_2^2 ) \;\ \right] \;\ \right\}
\label{XV}
\eeqa
This expression can then be used in Eqs.~(\ref{ghvv}), (\ref{ghvv*})
and (\ref{ghv*v*2}) to evaluate the decay widths. We have verified
that if we use Eq.~(\ref{XV}) with the definitions of Eq.~(\ref{AVBV})
into Eq.~(\ref{ghvv}) for the decay into two real vector bosons we
recover the results of ref.~\cite{hagiwara}. However our
expressions extend those results for the off-shell case.

\ni
The decays $H {\rightarrow} \gamma Z$ and $H {\rightarrow \gamma\gamma}$
appear at tree-level, the corresponding form factors,
Eq.~(\ref{haz_coupling}) and Eq.~(\ref{haa_coupling}), are:
\beq
A_{BSM} \equiv  \frac{2 \pi M_W^2 \tan\theta_W}{\alpha} \left[ \,\
\frac{f_W}{\Lambda^2}
- \frac{f_B}{\Lambda^2} + 4 \left(
\frac{f_{BB}}{\Lambda^2} \sin\theta_W^2 - \frac{f_{WW}}{\Lambda^2}
\cos\theta_W^2 \right) \,\ \right]
\eeq
\beq
I_{BSM} \equiv \frac{8 \pi M_W^2 \sin\theta_W^2}{\alpha} 
\left(\frac{f_{BB}}{\Lambda^2} +
\frac{f_{WW}}{\Lambda^2} \right)
\eeq
With these variables it is possible to compute the various Higgs decay
widths, including the interference of the new terms with the Standard
Model, and allowing for decays to virtual gauge bosons.

\ni
In this model the Branching Ratios to $\gamma \gamma$ and $\gamma
Z$ increase and these decays may become dominant for some region of
parameters. For $H {\rightarrow} WW$ and $H {\rightarrow} ZZ$ the new
contributions may interfere constructively or destructively with the
Standard Model terms. 
In Fig.~(\ref{fig_br}) the width and
branching ratios of the Higgs as a function of its mass are displayed for
the Standard Model and with the new contributions where all the
non-zero $f_i$ are assumed equal and $f_i/\Lambda^2$ = 100 TeV${}^{-2}$.

\noindent

\setcounter{figure}{7}
\begin{figure}
\begin{tabular}{cc}
\mbox{\epsfig{file=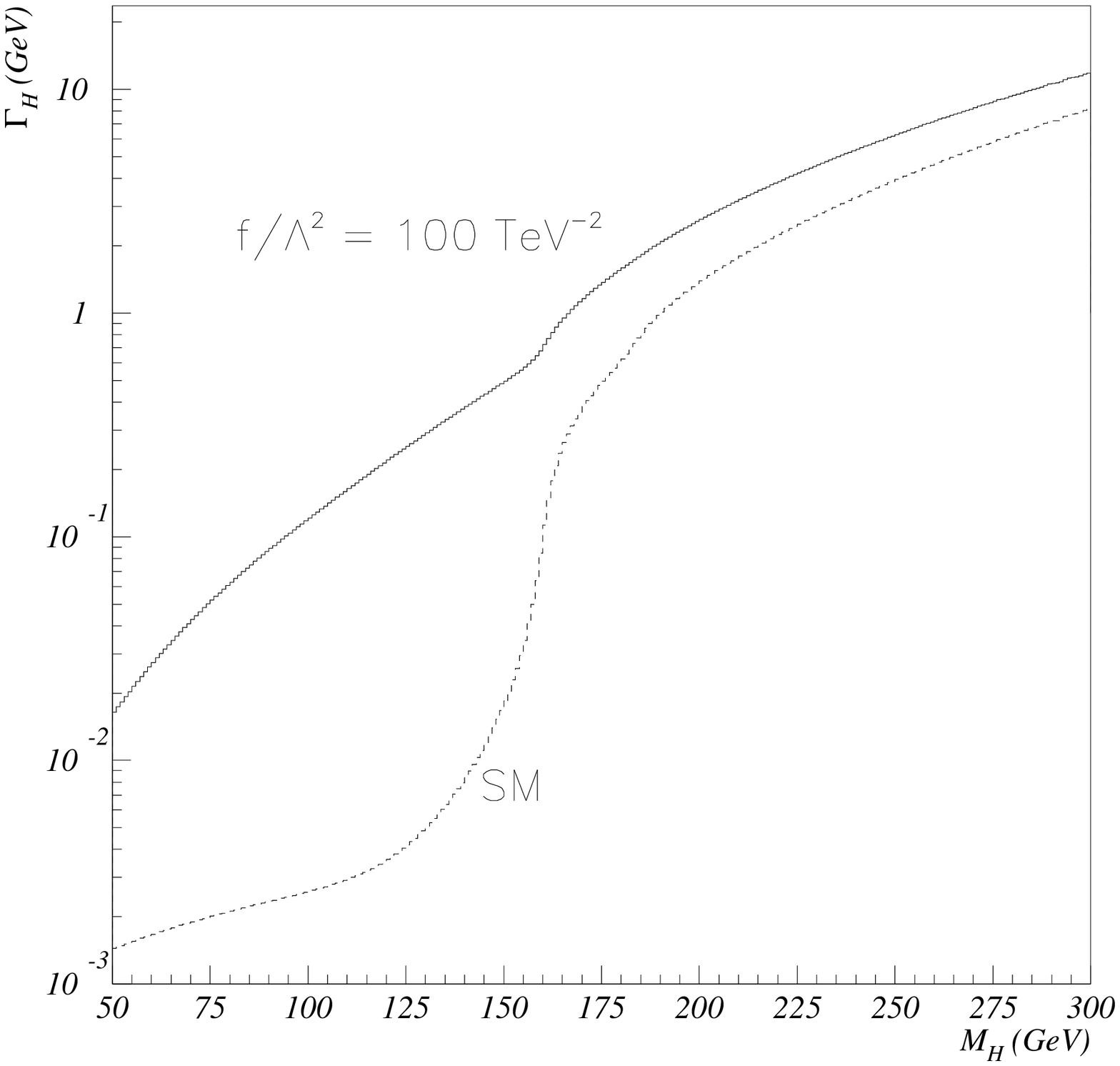,height=70mm,width=0.40\linewidth}}
&
\mbox{\epsfig{file=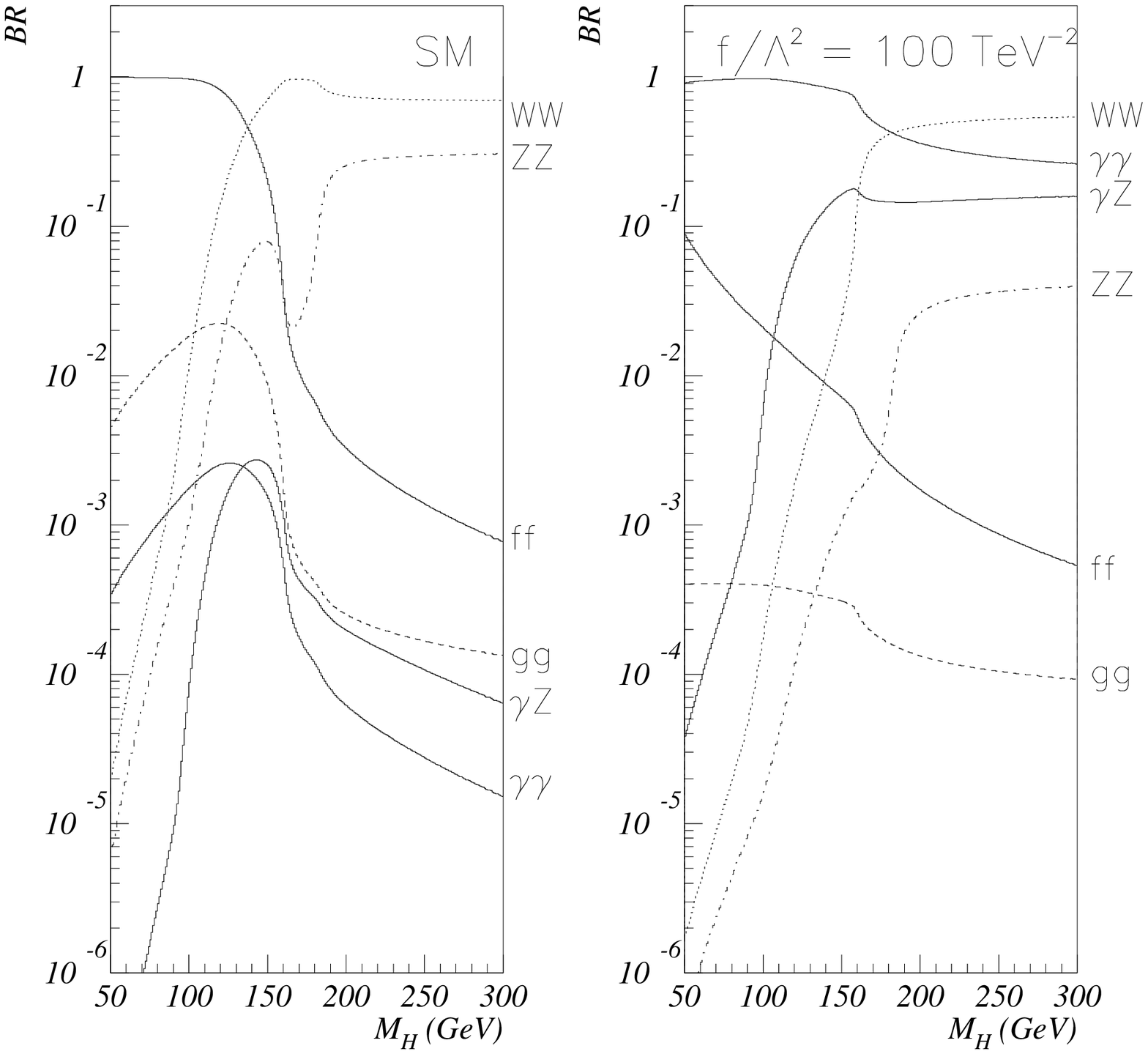,height=70mm,width=0.5\linewidth}}
\end{tabular}
\caption{Higgs Width and Branching Ratios as a function of its mass}
\label{fig_br}
\end{figure}

\begin{figure}
\begin{center}
\mbox{\epsfig{file=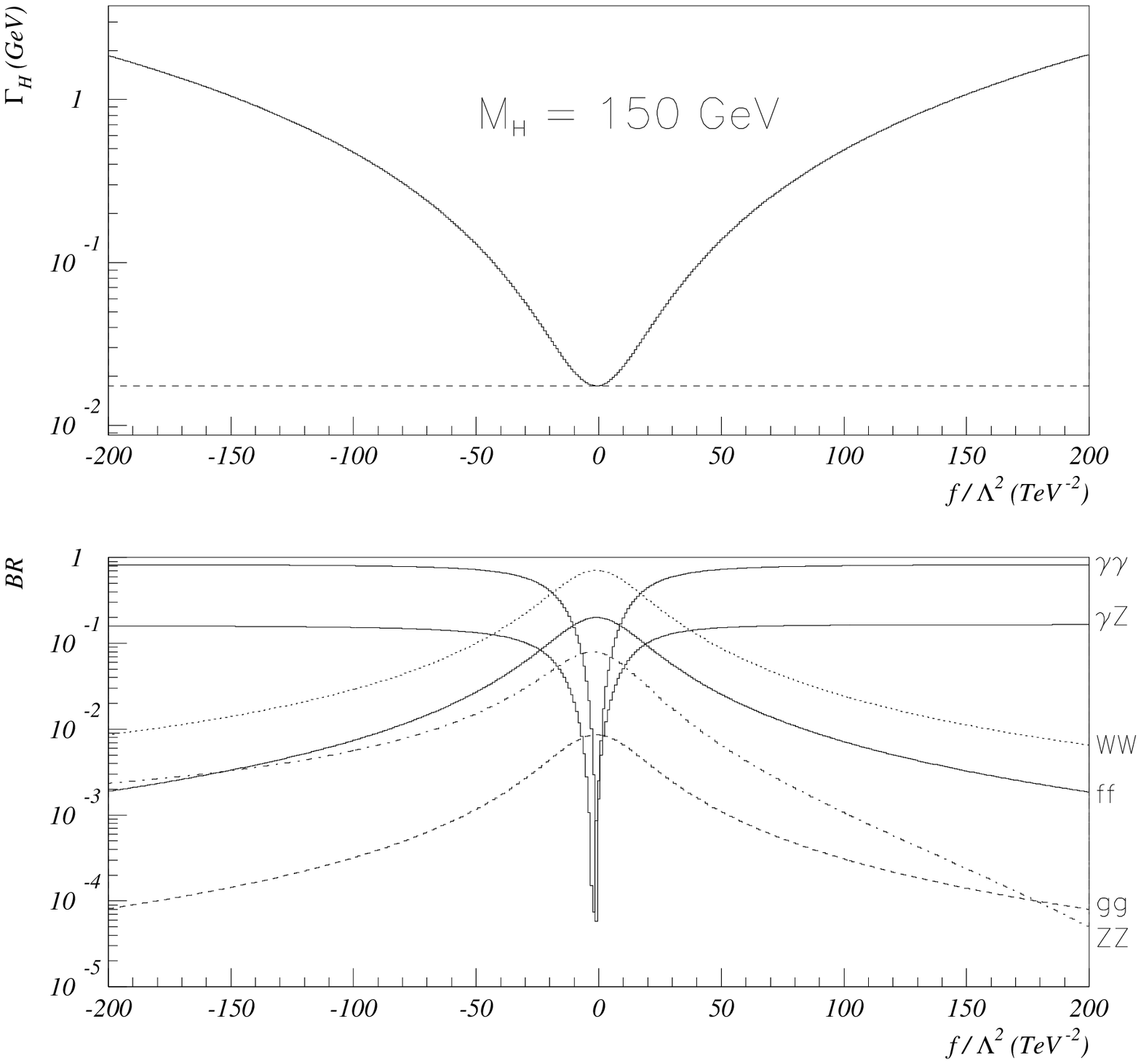,width=.7\linewidth}}
\end{center}
\caption{Higgs width as a function of $f/\Lambda^2$}
\label{fig_fvar}
\end{figure}

\begin{figure}
\begin{tabular}{cc}
\mbox{\epsfig{file=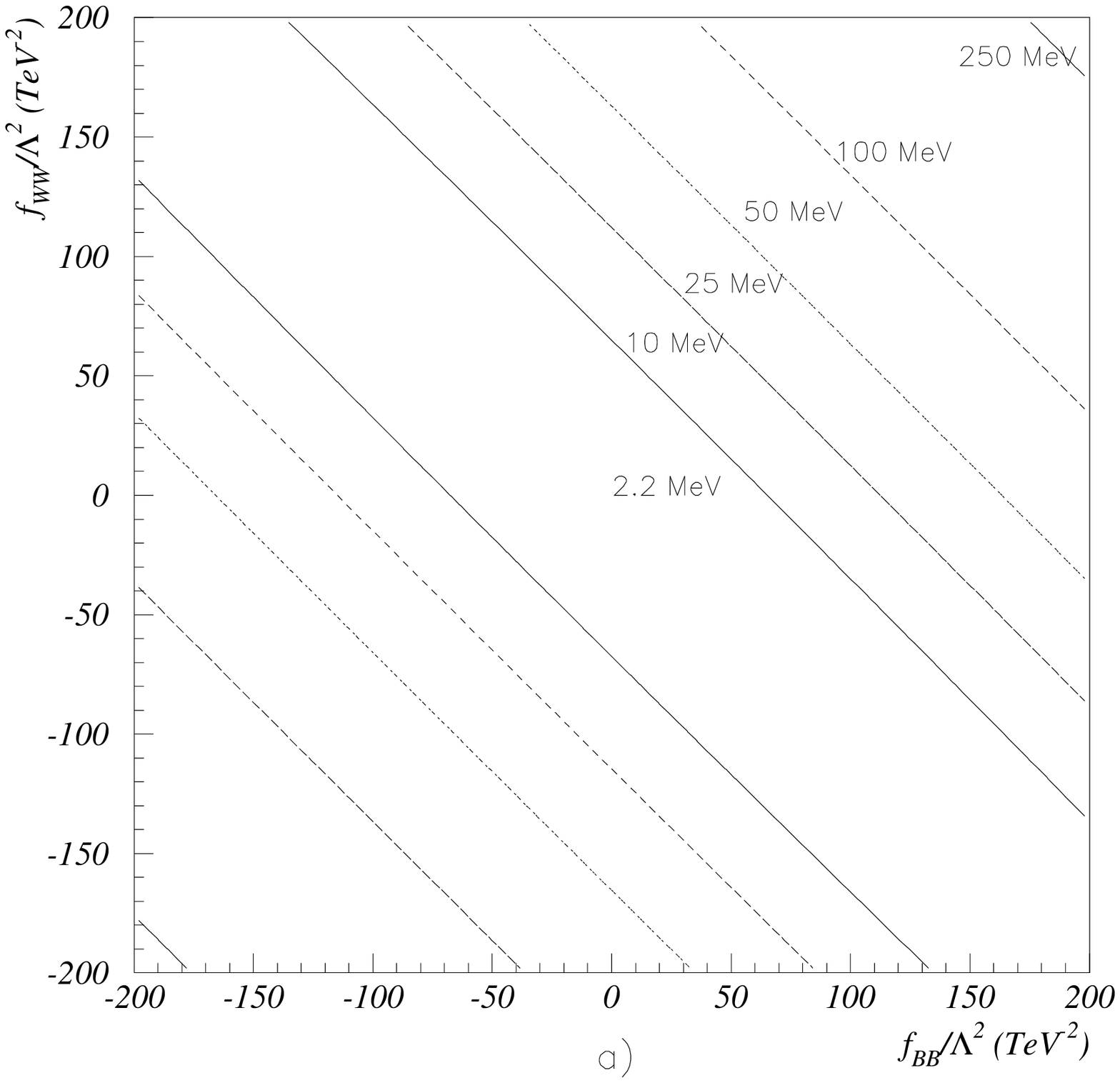,width=0.45\linewidth}}
&
\mbox{\epsfig{file=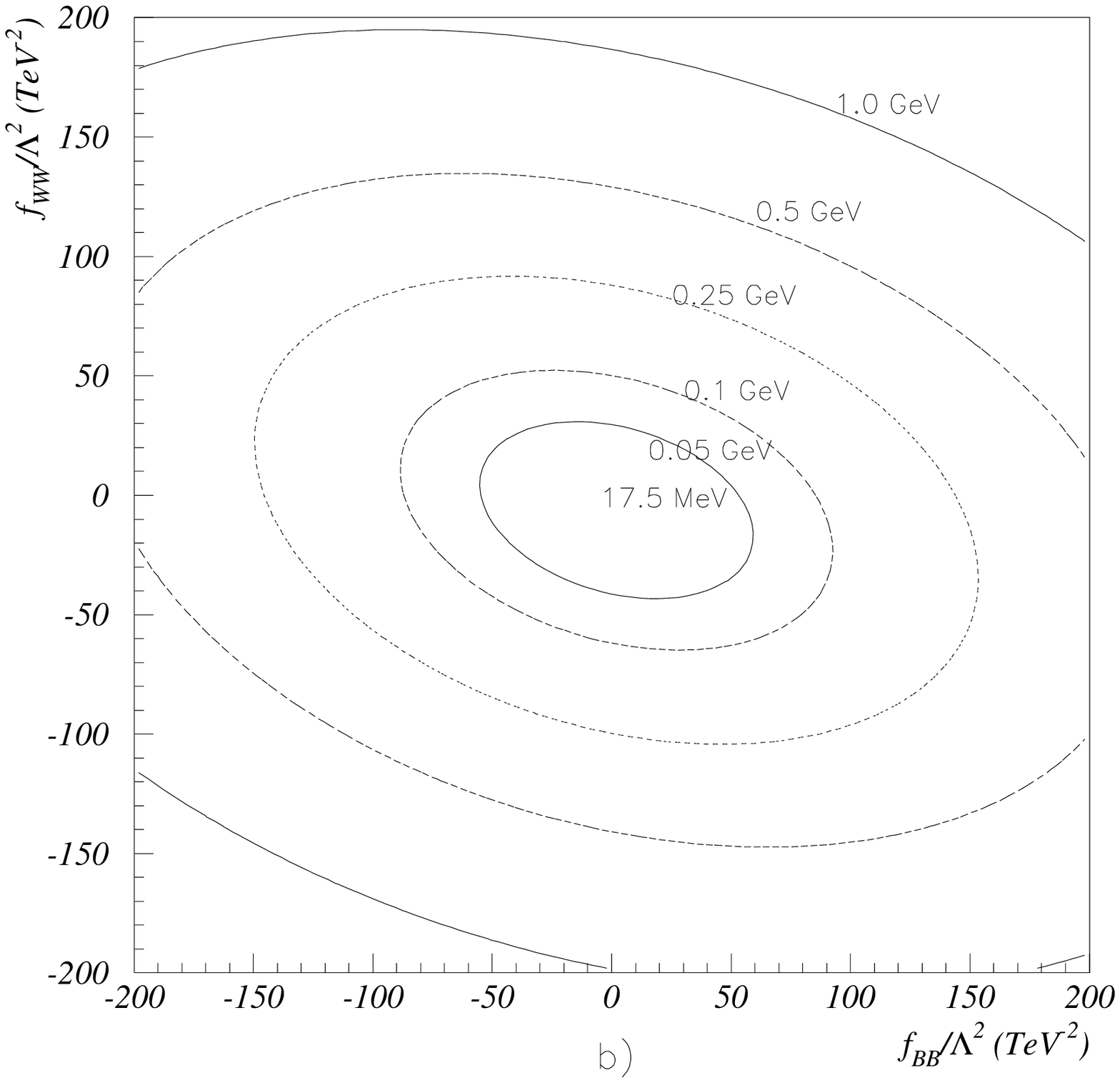,width=0.45\linewidth}}
\end{tabular}
\caption{Constant Higgs' Width lines as a function of $f_{BB}/\Lambda^2$
        and $f_{WW}/\Lambda^2$ for:
        a) $M_H = 85$ GeV and b) $M_H = 150$ GeV }
\label{fig_2d}
\end{figure}

\ni
The variation of the total width and branching ratios with $f/\Lambda^2$
is shown in Fig.~(\ref{fig_fvar}), for a Higgs boson mass of 150 GeV.
In Fig.~(\ref{fig_2d}) all $f_i$ except the ones contributing directly
to the H decay to $\gamma\gamma$ are set to 0. The variation with
$f_{BB}/\Lambda^2$ and $f_{WW}/\Lambda^2$ is displayed for two different
masses: 85 GeV and 150 GeV.

\section{Discussion}

In this paper we derive the complete set of formulas for the decay
widths of the Higgs boson in vector bosons. The formulas are valid
both for the Standard Model (SM) and for any arbitrary extension.
For the case of the decay into the $W^\pm$ and $Z^0$ the formulas are
also valid for off--shell decays. This is important for Higgs boson
masses close to the threshold of the production of one or two
real vector bosons. 
As many of these results have appeared before in the
literature~\cite{marciano,HHG,hagiwara2,hagiwara,ee500,djouadi}, sometimes for
particular cases, we will now comment on the comparison of our results
with those.

For the on--shell decay $H\ra VV$ our final expression Eq.~(\ref{ghvv}), is in
agreement with ref.~\cite{djouadi}. There is a factor 2 difference
with respect to ref.~\cite{ee500}. 
For the off-shell decay $H\ra VV^*$ our final expression, Eq.~(\ref{ghvv*3}) in
agreement with ref.~\cite{djouadi}. We are also in agreement with
Eq.~(6) of ref.~\cite{marciano} in the zero width limit. Eqs.~(9-10) of
ref.~\cite{marciano} are also consistent with our results and in
agreement with ref.~\cite{ee500}. 
For the off--shell decay $H \ra V^*V^*$ our result,
Eq.~(\ref{ghv*v*2}), is in agreement with ref.~\cite{djouadi} except for
the factor $\delta_V$. 
For the on--shell decay $H\ra \gamma \gamma$ we are in agreement with
refs.~\cite{HHG,hagiwara}, while for the on--shell decay $H\ra \gamma
Z$ we agree with ref.~\cite{HHG} but have a factor of two difference
with respect to ref.~\cite{hagiwara}. The formulas for the off-shell
decays $H\ra \gamma \gamma^*$ and $H\ra \gamma Z^*$ are either new, or
in agreement with refs.~\cite{barroso,babu}.

As our main contribution is to extend the formulas for an arbitrary
extension of the SM, including off-shell decays we studied, as an example,
the case of the gauge--invariant effective Lagrangian models of
ref.~\cite{hagiwara2,hagiwara}. Our expressions reproduce the results of
ref.~\cite{hagiwara} for the on--shell decays and extend them for the region
of the Higgs boson mass close to the two $W$'s threshold where the
off--shell decays have to be considered. This region is important for the
studies done at the Tevatron and at LEPII where these models have been
considered~\cite{novaes,LIP}.

\newpage

\appendix

\section*{Appendix A: Standard Model Feynman Rules}
\label{ApendiceA}

Because of the interference terms between the Standard Model (SM) and
possible extensions, it is important that we state our conventions for
the SM. We follow the conventions of ref.~\cite{HHG}. These differ in
some signs from the conventions used in refs.~\cite{barroso,babu}. For
the convenience of the reader we collect the most important Feynman
rules here.

\vskip 1cm
\beq
\hskip 1cm i\, g M_W\, g_{\mu \nu} \hskip 7cm i\, g M_Z \, g_{\mu \nu}
\eeq

\begin{picture}(0,0)
\put(0,-0.3){\psfig{figure=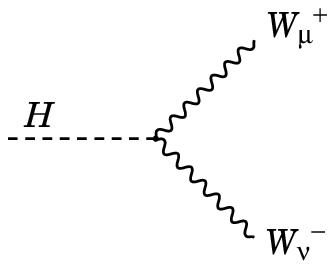}}
\end{picture}

\begin{picture}(0,0)
\put(8,0.7){\psfig{figure=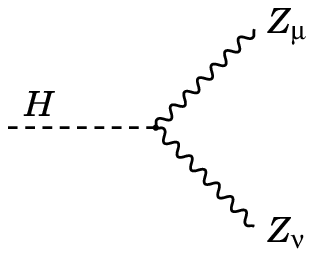}}
\end{picture}

\vskip 1cm
\beq
\hskip 1cm -i\, \frac{g\, m_f}{2 M_W}
\hskip 7cm -i\, e\, Q_f \gamma^{\mu}
\eeq

\begin{picture}(0,0)
\put(0,-0.3){\psfig{figure=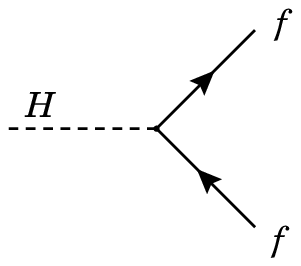}}
\end{picture}

\begin{picture}(0,0)
\put(8,0.7){\psfig{figure=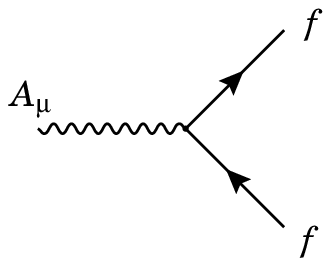}}
\end{picture}

\vskip 1cm
\beq
\hskip 3cm -i\, \frac{g}{\cos \theta_W} \gamma^{\mu} \left( g_V^f -
g_A^f \gamma_5 \right)
\eeq

\begin{picture}(0,0)
\put(3,-0.3){\psfig{figure=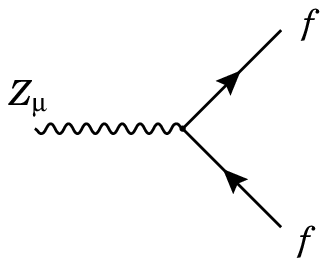}}
\end{picture}

\vskip 1cm
\beq
\hbox{\hskip 4cm} i\, g_V \left[\vb{14} g^{\mu \nu} (p_2 - p_3)^{\rho} +
g^{\nu \rho} (p_3 - p_1)^{\mu} +g^{\rho \mu} (p_1 - p_2)^{\nu} \right]
\eeq

\begin{picture}(0,0)
\put(0.5,-0.1){\psfig{figure=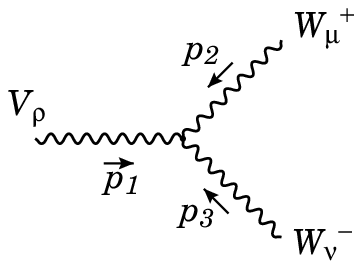}}
\end{picture}

\ni
for $V=A,Z$ with $g_A=e$, $g_Z=g \cos \theta_W$ and
\beq
g_V^f=\frac{1}{2}\, T_3^f -Q_f \sin^2 \theta_W 
\quad ; \quad
g_A^f=\frac{1}{2}\, T_3^f 
\eeq
where $Q_f$ is the charge of fermion $f$ in units of $|e|$.

\section*{Appendix B: The $J_1$ and $J_2$ functions}
\label{ApendiceB}

The explicit expressions for the functions $J_1$ and $J_2$ introduced
in Section~\ref{SecHGZ}, are~\cite{barroso,babu}
\beqa
J_1(q^2,M_H^2,M_X^2)&=&-  M_W^2\ 
C_0(q^2,0,M_H^2,M_X^2,M_X^2,M_X^2) \cr
&&\cr
J_2(q^2,M_H^2,M_X^2)&=&
\frac{1}{2} \frac{M_X^2}{q^2-M_H^2} \left[ \vbox to 18pt{}1 + 2 M_X^2\
C_0(q^2,0,M_H^2,M_X^2,M_X^2,M_X^2) \right. \cr
&& \cr
&&\left. + \frac{q^2}{q^2-M_H^2}
\left( B_0(q^2,M_X^2,M_X^2)- B_0(M_H^2,M_X^2,M_X^2)\right)\right]
\eeqa
where $B_0$ and $C_0$ are the Passarino-Veltman
functions\cite{thvelt} and $M_X$ is the mass of the particle in the
loop. These functions are related to the functions
$I_1$ and $I_2$ of ref.~\cite{HHG} by the following relations
\beqa
J_1(q^2,M_H^2,M_X^2)&=&I_2(\tau_X,\lambda_X)\cr
\vb{20}
J_2(q^2,M_H^2,M_X^2)&=&\frac{1}{4}\ I_1(\tau_X,\lambda_X)
\eeqa
with
\beq
\tau_X=\frac{4 M_X^2}{M_H^2}
\quad ; \quad
\lambda_X=\frac{4 M_X^2}{q^2}
\eeq
With these relations it is easy to verify that Eq.~(\ref{A_SM}) is in
agreement with Eq.~(2.22) of ref.~\cite{HHG}. To verify the
equivalence of Eq.~(\ref{I_SM}) with Eqs.~(2.16) and (2.17) of
ref.~\cite{HHG} for the on--shell decay $H\ra \gamma \gamma$
one has to note that 
\beqa
J_1(0,M_H^2,M_X)&=&I_2(\tau_X,\infty)=\frac{\tau_X}{2}\ f(\tau_X)\cr
\vb{20}
J_2(0,M_H^2,M_X)&=&\frac{1}{4}\ I_1(\tau_X,\infty)=-\frac{\tau_X}{8}
+\frac{\tau_X^2}{8}\ f(\tau_X)
\eeqa
where $f(\tau)$ is defined in Eq.~(2.19) of ref.~\cite{HHG}. Then we
get for the $W$ contribution
\beqa
I_{W}&=&-4 \left[\vbox to 18pt{}
-4 J_1(0,M_H^2,M_W^2) 
+ \left(6 + \frac{M_H^2}{M_W^2}\right)\ J_2(0,M_H^2,M_W^2)
\right]\cr
\vb{20}
&=& 16\, J_1(0,M_H^2,M_W^2) - \left(24 + \frac{16}{\tau_W}\right)\, 
J_2(0,M_H^2,M_W^2)\cr
\vb{20}
&=&
2 + 3 \tau_W + 3 \tau_W (2 -\tau_W)\, f(\tau_W)
\eeqa
and for a fermion of charge $Q_f$
\beqa
I_{F}&=&4\, Q_f^2 \left[\vbox to 18pt{}
- J_1(0,M_H^2,M_W^2) 
+ 4 J_2(0,M_H^2,M_W^2)
\right]\cr
\vb{20}
&=& Q_f^2 \left[
- 2\, \tau_f f(\tau_f) -2\, \tau_f + 2\tau_f^2 f(\tau_f) \right]\cr
\vb{20}
&=&
Q_f^2 \left[
-2\, \tau_f\left(\vb{16} 1 +(1 - \tau_f)\, f(\tau_f) \right) \right]
\eeqa
in agreement with Eq.~(2.17) of ref.~\cite{HHG}.

\end{document}